\documentclass[prl,amssymb,twocolumn,showpacs]{revtex4}

\usepackage{graphicx}% Include figure files

\begin{document}
\title{Topological Superconductivity and Superfluidity}
\author{Xiao-Liang Qi, Taylor L. Hughes, Srinivas Raghu and Shou-Cheng Zhang}

\affiliation{Department of Physics, McCullough Building, Stanford
University, Stanford, CA 94305-4045}

\begin{abstract}
We construct time reversal invariant topological superconductors and
superfluids in two and three dimensions which are analogous to the
recently discovered quantum spin Hall and three-d $Z_2$ topological
insulators respectively. These states have a full pairing gap in the
bulk, gapless counter-propagating Majorana states at the boundary,
and a pair of Majorana zero modes associated with each vortex. We
show that the time reversal symmetry naturally emerges as a
supersymmetry, which changes the parity of the fermion number
associated with each time-reversal invariant vortex. In the presence
of external T-breaking fields, non-local topological correlation is
established among these fields, which is an experimentally
observable manifestation of the emergent supersymmetry.
\end{abstract}

\maketitle

The search for topological states of quantum matter has become an
active and exciting pursuit in condensed matter physics. The quantum
Hall (QH) effect \cite{prange1990} provides the first example of a
topologically non-trivial state of matter, where the quantized Hall
conductance is a topological invariant\cite{thouless1982}. Recently,
the quantum spin Hall (QSH) state\cite{kane2005a,bernevig2006a} has
been theoretically predicted \cite{bernevig2006b} and experimentally
observed in HgTe quantum well systems\cite{koenig2007}. The time
reversal invariant (TRI) QSH state is characterized by a bulk gap, a
$Z_2$ topological number \cite{kane2005b}, and gapless helical edge
states, where time-reversed partners
counter-propagate\cite{wu2006,xu2006}.

Chiral superconductors in a time reversal symmetry breaking (TRB)
$(p_x+ip_y)$ pairing state in 2d have a sharp topological
distinction between the strong and weak pairing regimes
\cite{read2000}. In the weak pairing regime, the system has a full
bulk gap and gapless chiral Majorana states at the edge, which are
topologically protected. Moreover, a Majorana zero mode is trapped
in each vortex core\cite{read2000}, which leads to a ground state
degeneracy of $2^{n-1}$ in the presence of $2n$ vortices. When the
vortices wind around each other a non-Abelian Berry phase is
generated in the $2^{n-1}$ dimensional ground state manifold, which
implies non-Abelian statistics for the vortices\cite{ivanov2001}.
Chiral superconductors are analogous to the QH state--- they both
break time reversal (TR) and have chiral edge states with linear
dispersion. However, the edge states of a chiral superconductor have
only half the degrees of freedom compared to the QH state, since the
negative energy quasi-particle operators on the edge of a chiral
superconductor describe the same excitations as the positive energy
ones.

\begin{figure}[t!]
\begin{center}
\includegraphics[width=3in] {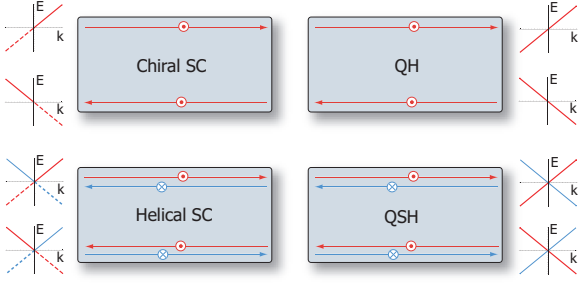}
\end{center}
\caption{(Top row) Schematic comparison of $2d$ chiral
superconductor and the QH state. In both systems, TR symmetry is
broken and the edge states carry a definite chirality. (Bottom row)
Schematic comparison of $2d$ TRI topological superconductor and the
QSH insulator. Both systems preserve TR symmetry and have a helical
pair of edge states, where opposite spin states counter-propagate.
The dashed lines show that the edge states of the superconductors
are Majorana fermions so that the $E<0$ part of the quasi-particle
spectra are redundant. In terms of the edge state degrees of
freedom, we have ${\rm (QSH)}={\rm (QH)}^2={\rm (Helical~SC)}^2={\rm
(Chiral~SC)}^4$.
 } \label{edgedispersion}
\end{figure}

Given the analogy between the chiral superconducting state and the
QH state, and with the recent discovery of the TRI QSH state, it
is natural to generalize the chiral pairing state to the helical
pairing state, where fermions with up spins are paired in the
$(p_x+ip_y)$ state, while fermions with down spins are paired in
the $(p_x-ip_y)$ state. Such a TRI state have a full gap in the
bulk, and counter-propagating helical Majorana states at the edge
(in contrast, the edge states of the TRI topological insulator are
helical Dirac fermions). Just as in the case of the QSH state, a
mass term for an odd number of pairs of helical Majorana states is
forbidden by TR symmetry, and therefore, a topologically protected
superconducting or superfluid state can exist in the presence of
time-reversal symmetry. Recently, a $Z_2$ classification of the
topological superconductor has been discussed in Refs
\cite{roy2006c,roy2008,schnyder2008}, by noting the similarity
between the Bogoliubov-de Gennes (BdG) superconductor Hamiltonian
and the QSH insulator Hamiltonian. The four types of topological
states of matter discussed here are summarized in Fig.
\ref{edgedispersion}. In this work, we give a $Z_2$ classification
of both the 2D and 3D cases which has a profound physical
implication. In two dimensions, we show that a time-reversal
invariant topological defect of a $Z_2$ non-trivial superconductor
carries a Kramers' pair of Majorana fermions. Let $N_F$ be the
operator which measures the number of fermions of a general
system, then the fermion-number parity operator is given by
$(-1)^{N_F}$. This operator is also referred to as the Witten
index\cite{witten1982},  which plays a crucial role in
supersymmetric theories. We prove the remarkable fact that in the
presence of a topological defect, the TR operator $\cal T$ changes
the fermion number parity, ${\cal T}^{-1} (-1)^{N_F} {\cal T} = -
(-1)^{N_F}$ locally around the defect in the $Z_2$ non-trivial
state, while it preserves the fermion number parity, ${\cal
T}^{-1} (-1)^{N_F} {\cal T} = (-1)^{N_F}$, in the $Z_2$ trivial
state. This fact gives a precise definition of the $Z_2$
topological classification of any TRI superconductor state and is
generally valid in the presence of interactions and disorder. A
supersymmetric operation can be defined as an operation which
changes the fermion number parity; therefore, in this precise
sense, we show that the TR symmetry emerges as a supersymmetry in
topological superconductors. Though supersymmetry has been studied
extensively in high energy physics, it has not yet been observed
in Nature. Our proposal offers the opportunity to experimentally
observe supersymmetry in condensed matter systems without any fine
tuning of microscopic parameters. The physical consequences of
such a supersymmetry is also studied.

As the starting point, we consider a TRI $p$-wave superconductor
with spin triplet pairing, which has the following $4\times 4$ BdG
Hamiltonian:
\begin{eqnarray}
H=\frac12\int d^2x\Psi^\dagger(x)
\left(\begin{array}{cc}\epsilon_{\bf
p}\mathbb{I}&i\sigma_2\sigma_\alpha\Delta^{\alpha
j}p_j\\h.c.&-\epsilon_{\bf
p}\mathbb{I}\end{array}\right)\Psi(x)\label{Hdouble}
\end{eqnarray}
with $\Psi(x)= \left(c_\uparrow(x),c_\downarrow(x),
c_\uparrow^\dagger(x),c_\downarrow^\dagger(x)\right)^T$,
$\epsilon_{\bf p}={\bf p}^2/2m-\mu$ the kinetic energy and
chemical potential terms and $h.c.\equiv
(i\sigma_2\sigma_\alpha\Delta^{\alpha j}p_j)^\dagger$. The TR
transformation  is defined as $c_\uparrow\rightarrow
c_\downarrow,~c_\downarrow\rightarrow -c_\uparrow$. It can be
shown that the Hamiltonian (\ref{Hdouble}) is time-reversal
invariant if $\Delta_{\alpha j}$ is a real matrix. To show the
existence of a topological state, consider the TRI mean-field
ansatz $\Delta^{\alpha 1}=\Delta(1,0,0),~\Delta^{\alpha
2}=\Delta(0,1,0)$. For such an ansatz the Hamiltonian
(\ref{Hdouble}) is block diagonal with only equal spin pairing:
\begin{eqnarray}
H=\frac12\int
d^2x\tilde{\Psi}^\dagger\left(\begin{array}{cccc}\epsilon_{\bf
p}&\Delta p_+&&\\
\Delta p_-&-\epsilon_{\bf p}&&\\
&&\epsilon_{\bf p}&-\Delta p_-\\
&&-\Delta p_+&-\epsilon_{\bf
p}\end{array}\right)\tilde{\Psi}\label{Hdoublepip}
\end{eqnarray}
with $\tilde{\Psi}(x)\equiv
\left(c_\uparrow(x),c_\uparrow^\dagger(x),
c_\downarrow(x),c_\downarrow^\dagger(x)\right)^T$, and
$p_\pm=p_x\pm ip_y$. From this Hamiltonian we see that the spin up
(down) electrons form $p_x+ip_y$ ($p_x-ip_y$) Cooper pairs,
respectively. In the weak pairing phase with $\mu>0$, the
$(p_x+ip_y)$ chiral superconductor is known to have chiral
Majorana edge states propagating on each boundary, described by
the Hamiltonian $H_{\rm edge}=\sum_{k_y\geq
0}v_Fk_y\psi_{-k_y}\psi_{k_y}$,%\label{Hpipedge}
where $\psi_{-k_y}=\psi_{k_y}^\dagger$ is the quasiparticle creation
operator \cite{read2000} and the boundary is taken to be parallel to
the $y$ direction. Thus we know that the edge states of the TRI
system described by Hamiltonian (\ref{Hdoublepip}) consist of spin
up and spin down quasi-particles with opposite chirality:
\begin{eqnarray}
H_{\rm edge}=\sum_{k_y\geq
0}v_Fk_y\left(\psi_{-k_y\uparrow}\psi_{k_y\uparrow}-\psi_{-k_y\downarrow}\psi_{k_y\downarrow}\right).\label{Hdoubleedge}
\end{eqnarray}
The quasi-particle operators
$\psi_{k_y\uparrow},~\psi_{k_y\downarrow}$ can be expressed in terms
of the eigenstates of the BdG Hamiltonian as
\begin{eqnarray}
\psi_{k_y\uparrow}&=&\int
d^2x\left(u_{k_y}(x)c_{\uparrow}(x)+v_{k_y}(x)c_{\uparrow}^\dagger(x)\right)\nonumber\\
\psi_{k_y\downarrow}&=&\int
d^2x\left(u_{-k_y}^*(x)c_{\downarrow}(x)+v_{-k_y}^*(x)c_{\downarrow}^\dagger(x)\right)
\end{eqnarray}
from which the time-reversal transformation of the quasiparticle
operators can be determined to be
${\cal{T}}^{-1}\psi_{k_y\uparrow}{\cal{T}}=\psi_{-k_y\downarrow},~{\cal{T}}^{-1}\psi_{k_y\downarrow}{\cal{T}}=-\psi_{-k_y\uparrow}$.
In other words, $(\psi_{k_y\uparrow},~\psi_{-k_y\downarrow})$
transforms as a Kramers' doublet, which forbids a gap in the edge
states due to mixing of the spin-up and spin-down modes when TR is
preserved. To see this explicitly, notice that the only
$k_y$-independent term that can be added to the edge Hamiltonian
(\ref{Hdoubleedge}) is
$im\sum_{k_y}\psi_{-k_y\uparrow}\psi_{k_y\downarrow}$ with $m\in
\mathbb{R}$. However, such a term is odd under TR, which implies
that any back scattering between the quasi-particles is forbidden by
TR symmetry. The discussion above is exactly parallel to the $Z_2$
topological characterization of the quantum spin Hall system. In
fact, the Hamiltonian (\ref{Hdoublepip}) has exactly the same form
as the four band effective Hamiltonian proposed in
Ref.\cite{bernevig2006b} to describe HgTe quantum wells with the QSH
effect. The edge states of the QSH insulators consist of an odd
number of Kramers' pairs, which remain gapless under any small
TR-invariant perturbation\cite{wu2006,xu2006}. A no-go theorem
states that such a ``helical liquid" with an odd number of Kramers'
pairs at the Fermi energy can not be realized in any bulk 1d system,
but can only appear as an edge theory of a 2d QSH
insulator\cite{wu2006}. Similarly, the edge state theory
(\ref{Hdoubleedge}) can be called a ``helical Majorana liquid",
which can only exist on the boundary of a $Z_2$ topological
superconductor. Once such a topological phase is established, it is
robust under any TRI perturbations.

The Hamiltonian (\ref{Hdouble}) can be easily generalized to three
dimensions, in which case $\Delta^{\alpha j}$ becomes a $3\times
3$ matrix with $\alpha=1,2,3$ and $j=x,y,z$. An example of such a
Hamiltonian is given by the well-known $^3$He BW phase, for which
the order parameter $\Delta^{\alpha j}$ is determined by an
orthogonal matrix $\Delta^{\alpha j}=\Delta u^{\alpha j}$,
$u\in{\rm SO(3)}$\cite{vollhardt1990}. Here and below we ignore
the dipole-dipole interaction term \cite{leggett1975} since it
does not affect any essential topological properties. By applying
a spin rotation, $\Delta^{\alpha j}$ can be diagonalized to
$\Delta^{\alpha j}=\Delta \delta^{\alpha j}$, in which case the
Hamiltonian (\ref{Hdouble}) has the same form as a 3d Dirac
Hamiltonian with momentum dependent mass $\epsilon({\bf p})={\bf
p}^2/2m-\mu$. We know that a band insulator described by the Dirac
Hamiltonian is a 3d $Z_2$ topological insulator for
$\mu>0$\cite{fu2007b,moore2007,roy2006b}, and has nontrivial
surface states. The corresponding superconductor Hamiltonian
describes a topological superconductor with 2d gapless Majorana
surface states. The surface theory can be written as
\begin{eqnarray}
H_{\rm surf}=\frac12\sum_{\bf k}v_F\psi_{-\bf
k}^T\left(\sigma_zk_x+\sigma_xk_y\right)\psi_{\bf k}
\end{eqnarray}
which remains gapless under any small TRI perturbation since the
only available mass term $m\sum_{\bf k}\psi_{-\bf
k}^T\sigma_y\psi_{\bf k}$ is time-reversal odd. We would like to
mention that the surface Andreev bound states in $^3$He-B phase
have been observed experimentally\cite{aoki2005}.

 To understand the physical consequences of the
nontrivial topology we study the TRI topological defects of the
topological superconductors. We start by considering the equal-spin
pairing system with BdG Hamiltonian (\ref{Hdoublepip}) in which spin
up and down electrons form $p_x+ip_y$ and $p_x-ip_y$ Cooper pairs,
respectively. A TRI topological defect can be defined as a vortex of
spin-up superfluid coexisting with an anti-vortex of spin-down
superfluid at the same position. In the generic Hamiltonian
(\ref{Hdouble}), such a vortex configuration is written as
$\Delta^{\alpha j}=[\exp\left(i\sigma_2\theta({\bf
r-r_0})\right)]^{\alpha j},~\alpha=1,2$ and $\Delta^{3j}=0$, where
$\theta({\bf r-r_0})$ is the angle of ${\bf r}$ with respect to the
vortex position ${\bf r}_0$. Since in the vortex core of a weak
pairing $p_x+ip_y$ superconductor there is a single Majorana zero
mode\cite{read2000,stone2006}, one immediately knows that a pair a
Majorana zero modes exist in the vortex core we study here. In terms
of the electron operators, the two Majorana fermion operators can be
written as
\begin{eqnarray}
\gamma_\uparrow&=&\int
d^2x\left(u_0(x)c_\uparrow(x)+u_0^*(x)c_\uparrow^\dagger(x)\right)\nonumber\\
\gamma_\downarrow&=&\int
d^2x\left(u_0^*(x)c_\uparrow(x)+u_0(x)c_\uparrow^\dagger(x)\right)
\end{eqnarray}
where we have used the fact that the spin-down zero mode wave
function can be obtained from the time-reversal transformation of
the spin-up one. The Majorana operators satisfy the
anti-commutation relation
$\left\{\gamma_\alpha,\gamma_\beta\right\}=2\delta_{\alpha\beta}$.
The TR transformation of the Majorana fermions is
\begin{eqnarray}
{\cal{T}}^{-1}\gamma_\uparrow
{\cal{T}}=\gamma_\downarrow,~{\cal{T}}^{-1}\gamma_\downarrow
{\cal{T}}=-\gamma_\uparrow .\label{TofMajorana}
\end{eqnarray}
Similar to the case of the edge states studied earlier, the
Majorana zero modes are robust under any small TRI perturbation,
since the only possible term $im\gamma_\uparrow\gamma_\downarrow$
which can lift the zero modes to finite energy is TR odd, {\it
i.e.},
${\cal{T}}^{-1}i\gamma_\uparrow\gamma_{\downarrow}{\cal{T}}=-i\gamma_\uparrow\gamma_{\downarrow}$.

The properties of such a topological defect appear identical to
that of a $\pi$-flux tube threading into a TRI topological
insulator\cite{qi2008,ran2008}, where a Kramers' pair of complex
fermions are trapped by the flux tube. However, there is an
essential difference. From the two Majorana zero modes
$\gamma_{\uparrow},\gamma_\downarrow$ a complex fermion operator
can be defined as
$a=\left(\gamma_\uparrow+i\gamma_\downarrow\right)/2$, which
satisfies the fermion anticommutation relation
$\left\{a,a^\dagger\right\}=1$. Since
$\gamma_\uparrow,\gamma_\downarrow$ are zero modes, we obtain
$\left[a,H\right]=0$ which implies that $a$ is the annihilation
operator of a zero-energy quasiparticle. Consequently, the ground
state of the system is at least two-fold degenerate, with two
states $|G_0\rangle$ and $|G_1\rangle=a^\dagger|G_0\rangle$
containing $0$ and $1$ $a$-fermions. Since $a^\dagger
a=\left(1+i\gamma_\uparrow \gamma_\downarrow\right)/2$, the states
$\left|G_{0(1)}\right\rangle$ are eigenstates of
$i\gamma_\uparrow\gamma_\downarrow$ with eigenvalues $-1(+1)$,
respectively. Thus from the oddness of
$i\gamma_\uparrow\gamma_\downarrow$ under TR we know that
$\left|G_0\right\rangle$ and $\left|G_1\right\rangle$ are
time-reversed partners. Note that superconductivity breaks the
charge $U(1)$ symmetry to $Z_2$, meaning that the fermion number
parity operator $(-1)^{N_F}$ is conserved. Thus, all the
eigenstates of the Hamiltonian can be classified by the value of
$(-1)^{N_F}$. If, say, $\left|G_0\right\rangle$ is a state with
$(-1)^{N_F}=1$, then
$\left|G_1\right\rangle=a^\dagger\left|G_0\right\rangle$ must
satisfy $(-1)^{N_F}=-1$. Since $\left|G_0\right\rangle$ and
$\left|G_1\right\rangle$ are time-reversal partners, we know that
in the Hilbert space of the zero-energy states the TR
transformation changes the fermion number parity:
\begin{eqnarray}
{\cal{T}}^{-1}(-1)^{N_F}{\cal{T}}=-(-1)^{N_F}.\label{TofNF}
\end{eqnarray}

Eq. (\ref{TofNF}) is the central result of this paper. At a first
glance it seems contradict the fundamental fact that the electron
number of the whole system is invariant under TR. Such a paradox
is resolved by noticing that there are always an even number of
topological defects in a closed system without boundary. Under the
TR transformation, the fermion number parity around each vortex
core is odd, but the total fermion number parity remains even as
expected. Once the anomalous transformation property (\ref{TofNF})
is established for a topological defect in a TRI superconductor,
it is robust under any TRI perturbation as long as the bulk
quasiparticle gap remains finite and other topological defects are
far away. Thus Eq. (\ref{TofNF}) is a generic definition of TRI
topological superconductors:

\begin{itemize}\item{\textsc{Definition I.} A two-dimensional
TRI superconductor is $Z_2$ nontrivial if and only if fermion
number parity around a TRI topological defect is odd under
TR.}\end{itemize}

A transformation changing fermion number by an odd number is  a
``supersymmetry"; thus, the TR symmetry emerges as a discrete
supersymmetry for each TRI topological defect. The same analysis
applies to the edge theory (\ref{Hdoubleedge}), which shows that
in the 1d helical Majorana liquid is a theory with TR symmetry as
a discrete supersymmetry.

\begin{figure}[htpb]
    \begin{center}
        \includegraphics[width=2.8in]{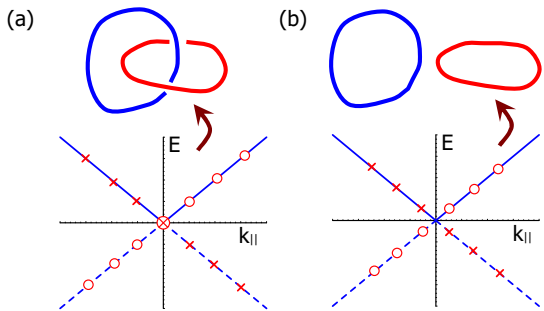}
    \end{center}
    \caption{Illustration of a 3d TRI topological superconductor with
    two TRI vortex rings which are (a) linked or (b) unlinked. The $E-k_\parallel$ dispersion relations show schematically the
    quasiparticle levels confined on the red vortex
    ring in both cases. ``$\circ$" and ``$\times$" stand for the quasiparticle levels that are Kramers' partners of each other.
    Only case  (a) has a pair of Majorana zero modes located on each vortex ring.}
    \label{fig:vortex3d}
\end{figure}

All the conclusions above can be generalized to 3d topological
superconductors. In the $^3$He BW phase the Goldstone manifold of
the order parameter is $\Delta^{\alpha j}=\Delta u^{\alpha j}\in
SO(3)\times U(1)$\cite{vollhardt1990,salomaa1987}. A time-reversal
invariant configuration satisfies $\Delta^{\alpha j}\in \mathbb{R}$,
which restricts the order parameter to $SO(3)$. Since
$\Pi_1(SO(3))=Z_2$, the TRI topological defects are 1d ``vortex"
rings. By solving the BdG equations in the presence of such vortex
rings, it can be shown that there are linearly dispersing
quasiparticles propagating on each vortex ring, similar to the edge
states of the 2d topological superconductor. However, for a ring
with finite length the quasi-particle spectrum is discrete.
Specifically, there may or may not be a pair of Majorana modes at
exactly zero energy. The existence of the Majorana zero modes on the
vortex rings turns out to be a topological property determined by
the linking number between different vortex rings. Due to the length
constraints of the present paper, we will write
 our conclusion and leave the details for a separate work: {\em
There are a pair of Majorana fermion zero modes confined on a
vortex ring if and only if the ring is linked to an odd number of
other vortex rings.} Such a condition is shown in Fig.
\ref{fig:vortex3d}. Consequently, the generalization of Definition
I to 3d is:

\begin{itemize}\item{\textsc{Definition II.} A 3d TRI superconductor is $Z_2$ nontrivial if and only if
the fermion number parity around one of the two mutually-linked
TRI vortex rings is odd under TR.}\end{itemize}

\begin{figure}[htpb]
    \begin{center}
        \includegraphics[width=3in]{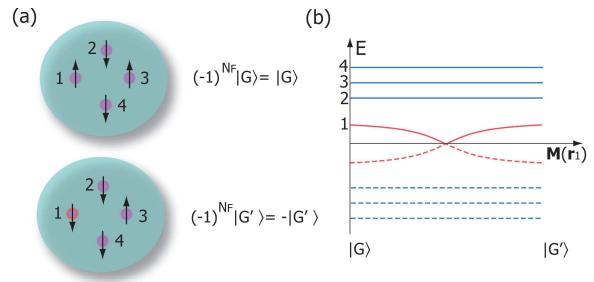}
    \end{center}
    \caption{(a) Illustration of a 2d TRI topological superconductor with four TRI topological
    defects coupled to a TR-breaking field. The arrows show the sign of the TR-breaking field ${\rm sgn}(M({\bf r}_s))$
    at each topological defect. In the two configurations shown, only the field around vortex 1 is flipped,
    leading to an opposite fermion number parity in the corresponding ground state $|G\rangle$ and $|G'\rangle$ (see text).
    (b) Illustration showing the flow of the energy levels when the upper configuration in figure (a) is deformed to
    the lower one. The flip of the TR-breaking field $M({\bf r}_1)$ leads to a level crossing at $M({\bf r}_1)=0$, where
    the fermion number
    parity in the ground state changes sign. }
    \label{fig:vortex2d}
\end{figure}

 Besides providing a
generic definition of the $Z_2$ topological superconductors, such
an emergent supersymmetry also leads to physical predictions.
Consider the 2d topological superconductor coupled to a weak
TR-breaking field $M({\bf r})$, which is classical but can have
thermal fluctuations. This situation can be realized in an
isolated superconductor with vortices pinned to quenched weak
magnetic impurities. The $n$-point correlation function of $M({\bf
r})$ can be obtained by
\begin{eqnarray}
\left\langle{\prod_{s=1}^nM({\bf r}_s)}\right\rangle\equiv\int
\frac{D[M({\bf r})]}Z\prod_{s=1}^nM({\bf r}_s){\rm
Tr}\left(e^{-\beta H[M]}\right)_{\rm even}\label{Mcorrelation}
\end{eqnarray}
in which the trace is restricted to states with an even number of
fermions. For a closed system with $N$ vortices, the leading order
effect of the TR-breaking field is to lift the degeneracy between
the two Majorana fermions in each vortex core. Consequently, the
perturbed Hamiltonian $H[M({\bf r})]$ to  first order can be
written as
\begin{eqnarray}
H[M({\bf r})]=\sum_{s=1}^{N}iM({\bf
r}_s)a_s\gamma_{s\uparrow}\gamma_{s\downarrow}\label{HofM}
\end{eqnarray}
in which $\gamma_{s\uparrow(\downarrow)}$ are the Majorana fermion
operators, and $a_s\in \mathbb{R}$ depend on the details of the
perturbation. The important fact is that the mass term induced is
linear in $M({\bf r})$ at the defect position ${\bf r}_s$, since
$i\gamma_{s\uparrow}\gamma_{s\downarrow}$ is TR odd.

Since the superconductor has a full gap, naively one would expect
all the correlations of $M({\bf r})$ field to be short ranged.
However, for a system with $N$ topological defects the $N$-point
correlation function has a long range order when ${\bf
r}_s,s=1,2,..N$ are chosen to be the coordinates of the
topological defects. In other words, the correlation function
\begin{eqnarray}
\lim_{|{\bf r}_i-{\bf r}_j|\rightarrow \infty,~\forall
i,j}\left\langle{\prod_{s=1}^NM({\bf r}_s)}\right\rangle\neq
0,\label{Mtopcorrelation}
\end{eqnarray}
though all the $n$ point correlations in Eq. (\ref{Mcorrelation})
with $n<N$ remain short ranged. Physically, such a non-local
correlation can be understood by comparing two states $|G\rangle$
and $|G'\rangle$, which are the ground states of the systems with
the field configurations $\mathcal{M}\equiv(M({\bf r}_1),M({\bf
r}_2),...,M({\bf r}_N))$ and $\mathcal{M}'\equiv(-M({\bf
r}_1),M({\bf r}_2),...,M({\bf r}_N))$, respectively. From
Hamiltonian (\ref{HofM}) it can be seen that $|G\rangle$ and
$|G'\rangle$ have opposite fermion number parity, since the fermion
number parity around the first topological defect
$i\gamma_{1\uparrow}\gamma_{1\downarrow}$ is reversed while that of
all the other topological defects remains invariant, as shown in
Fig. \ref{fig:vortex2d}. Without loss of generality, we can assume
$(-1)^{N_F}$ is even for $|G\rangle$ and odd for $|G'\rangle$. Since
the whole system is required to have an even number of fermions, the
lowest energy state in the Hilbert space for the field configuration
$\mathcal{M}'$ is not $|G'\rangle$, but the lowest quasiparticle
excitation $a_{\rm min}^\dagger|G'\rangle$. Thus, the two field
configurations $\mathcal{M}$ and $\mathcal{M}'$ have different free
energies, which leads to the non-vanishing correlation function in
Eq. (\ref{Mtopcorrelation}). Even when the topological defects are
arbitrarily far away, the energy difference between the two
configurations remains finite, which shows the non-local topological
correlation. Similar non-local correlations can also be obtained for
a 3d TRI topological superconductor with linked vortex rings.

Acknowledgement.---We acknowledge helpful discussions with S. B.
Chung, A. L. Fetter, L. Fu, R. Roy and S. Ryu. This work is
supported by the NSF under grant numbers DMR-0342832, the US
Department of Energy, Office of Basic Energy Sciences under contract
DE-AC03-76SF00515, and the Stanford Institute for Theretical Physics (S.R.).

\bibliography{TSC}
\end{document}